\documentclass[12pt]{article}
\usepackage{braket}
\usepackage{amsmath}
\usepackage{amsfonts}
\usepackage{amsmath}
\usepackage{amssymb}
\usepackage[mathscr]{eucal}

\begin{document} 
\title{Geometry of the Shannon mutual information in continuum QFT} 
\author{David R. Junior and Luis E. Oxman \\ \\
Instituto de F\'{\i}sica, Universidade Federal Fluminense,\\  
Campus da Praia Vermelha, Niter\'oi, 24210-340, RJ, Brazil. }

\maketitle

\begin{abstract}

We analyze geometric terms and scaling properties of the Shannon mutual information in the continuum. This is done for a free massless scalar field theory in $d$-dimensions, in a coherent state reduced with respect to a general differentiable manifold.  As a by-product, we find an expression for the reduced probability density of finding a certain field on a ball.  We will also introduce and compute the Fisher information that this probability carries about the location of the observation region.
This is an interesting information measure that refers to points in physical space, although in relativistic QFT they are labels and not fluctuating quantum observables. 

\end{abstract}

\section{Introduction}

The study of geometric entropies in quantum field theories is aimed at measuring the correlations between  degrees of freedom that live in and out a given region $\Omega $ of the physical space $\mathbb{R}^d$. For instance, in $(3+1)$ dimensions, the entanglement entropy for a spherical ball displays an area law \cite{9,  BKLS}, in analogy with the entropy of a Kerr-Newman black hole which is proportional to the area of the event horizon \cite{10}. Moving from the initial methods, based on the real time formulation in a discretized space, to the Euclidean formulation in the continuum led to new insights. 
It was discovered that important information is contained in the divergent part of the entanglement entropy 
 $S(\Omega)$  \cite{6}, \cite{2}, \cite{24}. As discussed in 
Ref. \cite{6}, in any Quantum Field Theory (QFT), its expected behavior is, 
\begin{equation}
S(\Omega) = g_{d-1}[\partial \Omega]\, \epsilon^{-(d-1)}+\dots + g_{1}[\partial \Omega]\, \epsilon^{-1} + g_0[\partial \Omega] \ln(\epsilon) + S_0(\Omega).
\label{arealaw} 
\end{equation} 
The coefficients $g_i$ are proportional to the $(i-1)$-th power of some characteristic length scale of $\Omega$, $\epsilon$ is a regulator, and $S_0$ is the finite part. The largest power of $\epsilon$ is multiplied by $g_{d-1}$, which has dimension of $[{\rm length}]^{d-1}$ thus corresponding to an area law. Terms proportional to $g_i$, with $i>0$, are not universal,  they depend on the regularization.  
However, the coefficient of the logarithmic divergence ($g_0$) is believed to be universal, giving important information about the conformal fixed points. This has been confirmed in $1+1$ dimensions, where $g_0 = \frac{c}{3}$ and $c$ is the central charge of the theory \cite{Holzhey, Cardy}. The search for new connections of this type is among the motivations to continue exploring different entropies in QFT.

More recently, other physical quantities like the ``R\'enyi-n" entropy \cite{8} and the Shannon mutual information \cite{22,7} have also been studied.  They are useful for the identification of universality classes at criticality \cite{6},\cite{7}. 
For R\'enyi-n, it is believed that in $3+1$ and $5+1$ dimensions $g_0$ is related to the coefficients of the Weyl anomaly \cite{8}. 
On the other hand, few results are known for the Shannon mutual information, specially in the continuum. Some of them were obtained in Ref. \cite{22, 7} for $d=1$.

In this work, we shall follow a real time formulation in the continuum, which could also be adapted for other entropies. In Ref. \cite{6}, such formulation was recognized as a previous step to understand the entropies associated with gauge fields. Initially, we will focus on how to compute reduced probabilities and the important role played by the nonlocal Poisson problem. Next, we discuss the Shannon mutual information for a massless scalar field in $d$ spatial dimensions, considering a region $\Omega$ whose border is a general differentiable manifold.  In the one dimensional case,  we will make contact with Ref. \cite{22}, while in higher dimensions we will obtain an expression for the geometric and divergent parts, as well as
the logarithmic dependence on the system size for two and three dimensional balls. Along this path, various results will be obtained. A relation between the Shannon and classical mutual correlations will be given, implying that the associated logarithmic divergences and scaling properties are simply related. In addition, the mutual information for a general coherent state and the vacuum will be shown to be equal, a property that is known to be valid for the entanglement entropy \cite{37}. We will also analyze the Fisher information 
as an interesting measure in Quantum Field Theory. For this aim, we will compute the information that the reduced probability carries about the location of the observation region. In this manner, we will get a space-dependent information in relativistic QFT, where the space points are field labels and not fluctuating quantum observables.

In sections \ref{vac} and \ref{determi}, we discuss the Poisson problem for the fractional Laplacian and its connection with reduced probabilities in a free massless scalar field theory. In particular, we obtain the reduced probability for a ball and a simple representation for the Shannon mutual information in the continuum. In section \ref{probe}, we introduce the Fisher information in QFT and compute it for a coherent field state. In section \ref{geomandiv}, we compute geometric and divergent terms. There, the scaling properties of the Shannon mutual information are also used to compute the logarithmic dependence on the system size in the case of a $d$-dimensional ball. Finally, in section \ref{conc}, we present our conclusions.

\section{The functional $\langle \phi| 0\rangle$ and the $(-\Delta)^{\frac{1}{2}}$ operator}
\label{vac}
 
 In a state described by a density matrix $\rho$, the geometric entanglement entropy between the degrees of freedom that live inside and outside a spatial region $\Omega \subset \mathbb{R}^d$ is given by,
\begin{equation}
S(\Omega) = -{\rm Tr} (\rho_{\Omega} \ln(\rho_\Omega))= -{\rm Tr} (\rho_{\bar{\Omega}} \ln(\rho_{\bar{\Omega}})) \;.
\end{equation}
Here, $\rho_\Omega$ and $\rho_{\bar{\Omega}}$ are the reduced density matrices with respect to external and internal degrees of freedom, respectively (we shall measure physical quantities with respect to $\hbar$, $c$ and $k_B$). 
This entropy  has been studied and computed in many situations. The first calculations were done 
for the vacuum state in a massless free field theory \cite{9,BKLS}, replacing the system with Hamiltonian
\begin{equation}
\hat{H}=\int dx \, \frac{1}{2} \left[  \hat{\pi}^2(x) + \hat{\phi}(x)(-\nabla^2)\hat{\phi}(x) \right] 
\label{cont}
\end{equation}
by a discretized version on a lattice,
\[
\Hat{H}_{\rm latt} = \frac{1}{2} \sum_{i=1}^N \hat{\pi}_i^2 + \frac{1}{2} \sum_{i,j=1}^N \hat{\phi}_i K_{ij} \hat{\phi}_j \;.
\] 
Correspondingly, the vacuum wave functional associated with eq. (\ref{cont}),
\begin{equation}
\Psi^{(0)} [\phi] = {\cal N}\,  \exp{ \left( -\frac{1}{2}\int dx\, \phi(x) [O \phi ](x)\right)} 
\makebox[.5in]{,} O = (-\Delta)^{\frac{1}{2}}\;,
\label{groundposition} 
\end{equation}
which depends on a nonlocal operator,  
was replaced by the ground state wave function,
\[
\Psi^{(0)}  (\phi) = \left( \det \frac{W}{\pi} \right)^{\frac{1}{4}}  e^{-\frac{1}{2} \phi^T W \phi}  \makebox[.5in]{,} W = \sqrt{K} \;.
\]
In this manner, when $\Omega$ is a ball of radius $r$ in three dimensions, an area law  was obtained for the first time \cite{9}, 
\[
S=0.30\, (r/a)^2 
\]
($a$ is the lattice spacing).

Another useful quantity is the Shannon mutual information, which  measures the amount of information that can be obtained about one random variable by observing the other \cite{26}.  In a general context, it is defined by,
\begin{equation}
I(U;V) = \int_U du \int_V dv\, p(u,v) \ln \left( \frac{p(u,v)}{p(u) \, p(v)}  \right)  \makebox[.3in]{,} \int du dv\, p(u,v) =1 \;,
\end{equation}
\begin{equation}
p(u) = \int_V dv\, p(u,v)    \makebox[.5in]{,}  p(v) = \int_U dv\, p(u,v) \;,
\end{equation}
where $U, V$ is a pair of continuum random variables with joint probability $p(u,v)$. Unlike the entanglement and R\'enyi-n entropies, the mutual information is basis dependent. 
In the  ground state  of a free field theory, the field variables in momentum-space are independent. Then, upon  dividing them into two momentum-space regions, the associated Shannon mutual information is zero. On the other hand,  we will be interested in computing the mutual information between internal and external degrees of freedom with respect to some spatial region $\Omega$, a quantity that shall be denoted by $I(\Omega ; \bar{\Omega})$. In this case, the fluctuating variables are the field values $\phi(x)$ that live inside and outside a given region $\Omega$, which are correlated due to the nonlocal nature of the operator $(-\Delta)^{\frac{1}{2}}$ in eq. \eqref{groundposition}. 
 The probability distribution for the complete system is
\[
P[\phi]=\Psi^*[\phi]\Psi[\phi] \makebox[.3in]{,} \Psi[\phi] = \braket{\phi|\Psi} \makebox[.3in]{,}
\hat{\phi}(x)\ket{\phi}=\phi(x)\ket{\phi} \;. 
\]
As usual, we can expand the field on a basis of eigenfunctions, 
\[ 
\phi (x) = \sum_n a_n\, \phi_n(x) \makebox[.3in]{,} O \phi_n(x)=\omega_n \phi_n(x)  \makebox[.3in]{,} \int_{\bar{\Omega}} dx\, \phi_n(x) \phi_m(x) = \delta_{nm} \;,
\]
and because of the mass dimensions $[\phi]= \frac{d-1}{2}$,  $[\phi_n]= \frac{d}{2}$, $[a_n]= -\frac{1}{2}$,
define the path integral measure as,
\[
[D\phi] = \prod_n \sqrt{\frac{\mu}{\pi}} \, da_n  \makebox[.5in]{,} [\mu]=1 \;.
\] 
For the vacuum state $|0 \rangle$, $\Psi^{(0)}[\phi] = \braket{\phi|0}$, this implies a normalization constant,
\begin{equation} 
{\cal N} = \left[ \det (\mu^{-1} O) \right]^{\frac{1}{4}} \;,
\label{Anorm}
\end{equation} 
where the functional determinant is computed with Dirichlet boundary conditions at infinity.

As we are interested in quantities in position space, it is important to carefully discuss how to deal with the fractional differential operator $(-\Delta)^{\frac{1}{2}}$.  Various equivalent definitions can be used; depending on the calculation, some of them could be preferred.  For example, this pseudo-differential operator can be defined using the Fourier transform. However, the geometry in $x$-space is hard to describe in momentum-space. Another interesting definition is the one introduced by Caffarelli and Silvestre in Ref. \cite{Caffarelli}, based on local objects defined on an extended space,
\begin{equation}
[(-\Delta)^{\frac{1}{2}} \phi ](x) \equiv -\partial_\tau u(x, 0) \;,
\label{cafa}
\end{equation}
\begin{equation}
(\partial_\tau^2 + \Delta ) u (x, \tau)=0\;, \makebox[0.8in]{{\rm for} $\tau > 0$}    \makebox[.5in]{,} u(x,0) = \phi(x) \;.  
\label{sp}
\end{equation}
From a physical point of view, this definition can be understood in terms of the vacuum wave functional, as computed in Euclidean spacetime, 
\[
\braket{0|\phi} \propto \int [Du]_\phi\, e^{-\int dx \int_0^\infty d\tau\,  \frac{1}{2}\, \partial_\mu u\, \partial_\mu u } \;,
\]
where $[Du]_\phi$ path-integrates over fields $u$ defined on the half ($\tau >0$) Euclidean spacetime  with the boundary condition $u(x,0) = \phi(x)$. Indeed, this is a  Gaussian path-integral that can be computed using the saddle-point method, which corresponds to eq. (\ref{sp}). An integration by parts, discarding spatial surface terms, gives
\[
\braket{0|\phi} \propto e^{ -\frac{1}{2}\int dx\,  u(x,0) (-\partial_\tau u |_{\tau=0} )  } = e^{ -\frac{1}{2}\int dx\,  \phi(x) (-\partial_\tau u |_{\tau=0} )  }  \;.
\]
Thus, comparing with eq. \eqref{groundposition}, the meaning of the identification in eq. \eqref{cafa} becomes clear.
This localization in a $(d+1)$-dimensional space of nonlocal objects acting on $d$-dimensional physical space underlines the success of the Euclidean spacetime method to compute entanglement entropies in the continuum. Another useful representation for powers of the Laplacian is in terms of a kernel \cite{13},
\begin{equation}
[(-\Delta)^{\gamma}\phi](x)  \equiv {\cal C}_{2\gamma} \int dy  \, \frac{(\phi( x)-\phi( y))}{|x-y|^{2\gamma+d}} \makebox[.5in]{,} 
{\cal C}_{2\gamma} = \frac{2^{2\gamma-\frac{d}{2}}}{\pi^{\frac{d}{2}}}
\frac{\Gamma(\frac{2\gamma+d}{2})}{\Gamma{(-\gamma})} \;. 
\label{def-part}
\end{equation}
In this form, it becomes evident that, for the vacuum state in eq. (\ref{groundposition}), the probability density for any constant configuration $\phi(x) \equiv {\rm const.}$  is the same than that for $\phi(x) \equiv 0$. This is expected for the massless case, due to the field translation symmetry of the Lagrangian. This form is also suited to solve the Poisson problem (see Refs. \cite{13}, \cite{20} and \cite{Stinga}), 
\[
\big[(-\Delta)^{\gamma} \varphi \big](x) = 0  
\]
with $x$ on a spatial region $\Omega \in \mathbb{R}^d$. A characteristic of the nonlocal case, which will play a relevant role in what follows, is that in order for the solution to be unique it is necessary to specify conditions on the whole complement $\bar{\Omega}$ \cite{20}. For example, for a ball $B$ of radius $r$ centered at the origin (internal problem), the solution is given by\footnote{Field labels will denote the region where the boundary condition is satisfied.},     
\begin{equation}   
 \varphi_{\bar{B}} (x) = \left\{ \begin{array}{ll}
          f_{\bar{B}}(x) & \mbox{if $x \in \bar{B} $} \\
       \int_{\bar{B}} dy\, \mathscr{P}(y,x) f_{\bar{B}}(y) & \mbox{if $x \in B$},\end{array} \right. 
       \label{solu}
\end{equation} 
where $\mathscr{P}$ is the Poisson kernel, 
\begin{equation}
\mathscr{P}(y,x) = {\cal C}_{d,\gamma} \left(\frac{ r^2-|x|^2 }{|y|^2 -r^2}\right)^{\gamma}\frac{1}{|x-y|^d} \makebox[.5in]{,}
{\cal C}_{d,\gamma} = \frac{\Gamma(\frac{d}{2}) \sin(\pi \gamma)}{\pi^{\frac{d}{2}+1}} \;.
\label{Pik}
\end{equation}
For the external problem, the solution is,
\begin{equation}
\varphi_{B} (x)  = \left\{ \begin{array}{ll}
          f_{B}(x) & \mbox{if $x \in B $} \\
       \int_{B} dy\, \mathscr{P}(x,y) f_{B}(y) & \mbox{if $x \in \bar{B}$}.\end{array} \right. 
       \label{fiB}
\end{equation}  
Notice that the expressions \eqref{solu} and (\ref{fiB}) depend on the data $f_{\bar{B}}(y)$ and $f_B(y)$, on   
$\bar{B}$ and $B$, respectively. 
It is instructive to discuss  $\gamma\rightarrow 1$, to check that the solution becomes dependent only on data at the boundary $\partial B$ in this limit. Changing to hyperspherical coordinates in $y$, for $x \in \bar{B}$, we have,
\begin{equation} 
 \varphi_{B} (x)  = {\cal C}_{d,\gamma}\int_r^\infty |y|^{d-1} d|y|\int d\Omega_d \, \frac{(r^2-|x|^2)^\gamma f_\Omega(|y|,\Omega_d)}{|x-y|^d(|y|-r)^\gamma (|y|+r)^\gamma}  \;.
\end{equation}
Then, making $|y|-r\equiv \alpha$ and
\begin{equation}
\xi(x,r,\alpha)\equiv \int d\Omega_d \,  \frac{(\alpha+r)^{d-1}(r^2-|x|^2)^\gamma f_\Omega(\alpha,\Omega_d)}{(\alpha+2r)^\gamma} \;,
\end{equation} 
we get
\begin{equation}
\varphi_{B} (x)  =  \frac{\Gamma(\frac{d}{2}) \sin(\pi \gamma)}{\pi^{\frac{d}{2}+1}}\int_0^\infty d\alpha \, \frac{\xi(x,r,\alpha)}{\alpha^\gamma} \;,
\label{intk}
\end{equation}
which contains the distribution
\begin{equation}
\alpha^{-\gamma}_+  = \left\{ \begin{array}{ll}
          \alpha^{-\gamma} & \mbox{if $x >0 $} \\
      0 & \mbox{if $x < 0$}\;.\end{array} \right.
\end{equation} 
In the $\gamma \to 1$ limit, as this distribution has a simple pole at $\gamma = 1$ (see Ref. \cite{gelfand}), and $\sin(\pi \gamma) \to 0$, $\varphi_{B} (x) $ in eq. (\ref{intk}) only receives a contribution of the residue,
which is proportional to $\delta(\alpha)$. Therefore, we get,
\begin{eqnarray}
\lefteqn{\varphi_{B} (x) \to \frac{\Gamma(\frac{d}{2})\xi(x,r,0)}{\pi^{\frac{d}{2}}}  }
\nonumber \\
&& =
 \frac{\Gamma(\frac{d}{2}+1)}{2\pi^{\frac{d}{2}}}\int d\Omega_d \,\frac{ (|x|^2-r^2)r^{d-2} f_\Omega(r,\Omega_d)}{|x-y|^d}    \;,
\end{eqnarray}
which only depends on $y$ values with $|y|=r$, at the border of $B$. In particular, for $d=3$, using $\Gamma(\frac{3}{2})=\frac{1}{2}\sqrt{\pi}$, 
\begin{equation}
\int_{\bar{B}} dy\, \mathscr{P}(y,x) f_\Omega (y) = \frac{1}{4\pi}\int d\Omega_3\frac{(|x|^2-r^2)r \ f_\Omega(r,\Omega_3)}{|x-y|^3} \;,
\end{equation}
which is the usual solution for the electrostatic problem in the interior of a sphere, with the potential specified on its boundary. The representation of the fractional Laplacian as a singular integral is known in $\mathbb{R}^n$ and in a hyperbolic space $\mathbb{H}^n$ \cite{banica}, while the Poisson kernel is only  known when $\Omega$ is a ball. However, to obtain the mutual information, an explicit expression for these quantities will not be necessary.

\section{Mutual information in the continuum}
\label{determi}

\label{Reduced probability densities}
To compute the Shannon mutual information in QFT we must deal with reduced probability densities.
Namely, the probability distribution of measuring a field configuration on a given region $\Omega$ irrespective of the value on its complement $\bar{\Omega}$. In section \ref{vac}, we have seen that the local representation of $(-\Delta)^{\frac{1}{2}}$ in an extended space is in line with the Euclidean spacetime representation of the vacuum wave functional. Similarly, we will see that the fractional operator and its ensuing Poisson problem is in line with the determination of the reduced probability functionals. The very nonlocal nature of the fractional Poisson problem on $\Omega$, which requires a boundary condition on $\bar{\Omega}$ in order to have a unique solution, is the key ingredient to obtain the reduced probability by the saddle-point method. A condition on the whole region $\bar{\Omega}$, not only at its boundary, is precisely what we need in the path-integrals to define reduced probabilities. This is in contrast to what happens in usual calculations in QFT with boundaries, involving second order differential operators. For example, to compute probability amplitudes in Quantum Mechanics we have to choose initial and final conditions at the border of a time interval, accordingly, this uniquely determines the saddle point $x_c(t)$ , as it satisfies a second order differential equation. 

\subsection{Reduced probabilities}

Manipulations based on the nonlocal saddle-point are esssential to account for the correlations between in and out modes, which are expected to be manifested in the mutual information. 
The reduced probability density of getting $f_{\Omega}(x)$, $x \in \Omega$, when measuring the field $\phi$ on $\Omega$, irrespective of the field value on $\bar{\Omega}$, is given by,
\begin{equation}
P_\Omega[f_\Omega] = \int [D\phi]_{f_\Omega}\, P[\phi] \;.
\label{POm} 
\end{equation}
where the measure $[D\phi]_{f_\Omega}$ integrates over the fields $\phi(x)$ with the condition,
\begin{equation}
\phi(x)= f_{\Omega}(x) \makebox[.5in]{,}  x \in \Omega \;.
\end{equation}
For the vacuum state $\Psi^{(0)}[\phi]$ (cf. eqs. \eqref{groundposition}, (\ref{Anorm})), to integrate the probability density, 
\begin{equation}
P^{(0)}[\phi] = \left[ \det (\mu^{-1} O) \right]^{\frac{1}{2}} \,  \exp{ \left( -\int dx\, \phi(x) [O \phi ](x)\right)} \;,
\end{equation}
we should find the field that minimizes
\begin{equation}
\int dx\, \phi(x) [O\phi ](x) \;,
\end{equation}
with respect to variations $\phi(x) \to \phi(x) + \delta \phi (x)$ such that $\delta\phi(x) =0$, for $x \in \Omega$. That is, 
\begin{equation}
\int dx\, \left[\delta\phi(x) [(-\Delta)^{\frac{1}{2}}\phi ](x) + \phi(x) [(-\Delta)^{\frac{1}{2}} \delta\phi ](x) \right] =0 \;.
\end{equation}
Using the fractional Laplacian in eq. (\ref{def-part}), we can integrate by parts,
\begin{eqnarray}
\lefteqn{\int dx \, \phi(x)  [(-\Delta)^{1/2}\delta\phi ](x) =} 
\nonumber \\
&&  = {\cal C}_{1}\int dx \int dy  \left(\frac{\phi(x)\delta\phi(x)-\phi(x)\delta\phi(y)}{|x-y|^{d+1}}\right) 
\nonumber \\
&& = {\cal C}_{1}\int dx \int dy\,  \frac{\delta\phi(x)\phi(x)}{|x-y|^{d+1}} - {\cal C}_1\int dx \int dy\, \frac{\delta\phi(x)\phi(y)}{|y-x|^{d+1}} 
\nonumber \\ 
&& = \int dx \, \delta\phi(x) [(-\Delta)^{1/2}\phi](x) \;,
\nonumber 
\end{eqnarray}
to get the variation,
\begin{equation}
2\int_{\bar{\Omega}} dx\, \delta\phi(x) [(-\Delta)^{\frac{1}{2}}\phi ](x)   =0 \;.
\end{equation}
Then, we arrive at a unique saddle point, which satisfies the nonlocal Poisson problem, 
\begin{equation}
[O\varphi_\Omega](x) = 0   \makebox[.3in]{,} x\in  \bar{\Omega} 
\makebox[.5in] {,}
  \varphi_\Omega(x) = f_{\Omega}(x) \makebox[.3in]{,}  x \in \Omega\;.
\label{intui} 
\end{equation}
The change of variables $\phi (x) = \varphi_\Omega(x) + \varepsilon (x) $, leads to,
\begin{equation} 
P^{(0)}_\Omega[f_\Omega] =P^{(0)}[ \varphi_\Omega] \int [D\varepsilon]_{\bar{\Omega}}\,  \exp\left[-\int_{\bar{\Omega}} dx\, \varepsilon(x) [O
\varepsilon] (x)\right]  \;,  
\end{equation} 
where the measure $[D\varepsilon]_{\bar{\Omega}}$  integrates over the fields $\varepsilon(x)$ such that,
\begin{equation}
 \varepsilon(x)= 0 \makebox[.5in]{,}  x \in \bar{\Omega} \;.
\end{equation} 
As before, expanding,
\[ 
\varepsilon (x) = \sum_i b_i\, \varepsilon_i(x) \makebox[.3in]{,} O \varepsilon_i(x)=\omega_i \varepsilon_i(x)  \makebox[.3in]{,} \int_{\bar{\Omega}} dx\, \varepsilon_i(x) \varepsilon_j(x) = \delta_{ij} \;,
\]
and defining the path integral measure as,
\[
[D\varepsilon]_{\bar{\Omega}} = \prod_i \sqrt{\frac{\mu}{\pi}} \, da_i \;,
\] 
we get,
\begin{equation} 
P_\Omega[f_\Omega] = \left[{\det}_{\bar{\Omega}} (\mu^{-1} O)\right]^{-\frac{1}{2}} P[ \varphi_\Omega]  \;.
\end{equation}
Proceeding in a similar way with the probability density reduced with respect to degrees in $\Omega$, we obtain\footnote{Labels of functional determinants denote the region where the eigenvalue problem is defined.}, 
\begin{eqnarray}
P^{(0)}_\Omega[f_\Omega] & = & \frac{[\det (\mu^{-1} O)]^{1/2}}{[\det_{\bar{\Omega}} (\mu^{-1}O)]^{1/2}} \exp\left[- \int dx \, \varphi_\Omega (x)[O \varphi_\Omega] (x)\right]  \;,
\label{Om}
\end{eqnarray}
\begin{eqnarray} 
P^{(0)}_{\bar{\Omega}} [f_{\bar{\Omega}}] & = & \frac{[\det (\mu^{-1}O)]^{1/2}}{[\det_{\Omega} (\mu^{-1}O)]^{1/2}} \exp\left[- \int dx \, \varphi_{\bar{\Omega}} (x)[O \varphi_{\bar{\Omega}}] (x)\right]  \;,
\label{Omcom}
\end{eqnarray}
where $\det_{\bar{ \Omega}} O$ ($\det_{\Omega} O$) is the determinant with Dirichlet boundary conditions on $\Omega$ ($\bar{\Omega}$) and $ \varphi_{\bar{\Omega}} (x)$ satisfies (\ref{intui}) with $\Omega \leftrightarrow \bar{\Omega}$.   

\subsection{Shannon mutual information in the vacuum}

Now, consider,  the mutual information,
\begin{eqnarray} 
I(\Omega;\bar{\Omega}) = \int [D\phi]\, P[\phi] \ln \left( \frac{P[\phi]}{P_\Omega[f_\Omega] \, P_{\bar{\Omega}}[f_{\bar{\Omega}}]}  \right)  \;.
\label{IOm}
\end{eqnarray}
For the ground state, we have
\begin{eqnarray} 
\lefteqn{ I(\Omega;\bar{\Omega}) =
\ln \left[ \frac{(\det_\Omega (\mu^{-1}O))^{1/2} (\det_{\bar{\Omega}} (\mu^{-1} O))^{1/2}}{(\det (\mu^{-1}O))^{1/2}} \right]  } \nonumber \\
&& + \int [D\phi]\, P^{(0)}[\phi]   \int dx \, \left[  \varphi_\Omega (x)[O \varphi_\Omega] (x) +  \varphi_{\bar{\Omega}} (x)[O \varphi_{\bar{\Omega}}] (x) - \phi (x)[O \phi] (x)   \right] \;. \nonumber  
\label{} 
\end{eqnarray} 
To obtain an expression for the second line, we can proceed as follows. If at the very beginning of the calulation we introduce an adimensional constant $J$, replacing $O \to JO$, $P^{(0)}[\phi]$ and $P^{(0)}_\Omega[f_\Omega]$ get replaced by,
\[
  P^{(0)}_J[\phi] = \left[ \det (\mu^{-1} JO) \right]^{\frac{1}{2}}\,  \exp{ \left( -\int dx\, \phi(x) [JO\phi ](x)\right)}\;,
\]
\[
P^{(0)}_{\Omega,J}[f_\Omega]  =  \frac{[\det (J\mu^{-1} O)]^{1/2}}{[\det_{\bar{\Omega}} (J\mu^{-1}O)]^{1/2}} \exp\left[- J\int dx \, \varphi_\Omega (x)[O \varphi_\Omega] (x)\right]  \;, 
\] 
which are normalized,
\[
\int [D\phi]\, P^{(0)}_J[\phi] =1 \makebox[.5in]{,} \int [Df_\Omega]\, P^{(0)}_{\Omega,J}[f_\Omega] =1 \;.
\]
The second condition is obtained from the first by using the splitting,
\begin{equation}
\int [D\phi] = \int [Df_\Omega]  \int [D\phi]_{f_\Omega} \;.
\label{split}
\end{equation} 
We can take derivatives of the normalization conditions with respect to $J$, set $J=1$, and use,
\[
\left. \frac{d}{dJ}\, P^{(0)}_J[\phi] \right|_{J=1} =  \frac{1}{2}\, P^{(0)}[\phi]  \left. \frac{d}{dJ}\, \ln \det (\mu^{-1} JO) \right|_{J=1}   - P^{(0)}[\phi]\int dx\, \phi(x) [O\phi ](x) \;,
\]
\begin{eqnarray}
\lefteqn{ \left. \frac{d}{dJ}\, P^{(0)}_{\Omega,J}[f_\Omega] \right|_{J=1} = } \nonumber \\
&&  \frac{1}{2}\, P^{(0)}_{\Omega}[f_\Omega] \, \frac{d}{dJ}\, \ln \left[ \frac{\det (J\mu^{-1} O)}{\det_{\bar{\Omega}} 
(J\mu^{-1}O)} \right]_{J=1}   - P^{(0)}_{\Omega}[f_\Omega] \int dx \, \varphi_\Omega (x)[O \varphi_\Omega] (x)\;, \nonumber
\end{eqnarray} 
to obtain,
\[
 \int [D\phi]\, P^{(0)}[\phi]\int dx\, \phi(x) [O\phi ](x) =  \frac{1}{2} \left. \frac{d}{dJ}\, \ln \det (\mu^{-1} JO) \right|_{J=1}  \;,
\]
\[
 \int [Df_\Omega]\, P^{(0)}_{\Omega}[f_\Omega] \int dx \, \varphi_\Omega (x)[O \varphi_\Omega] (x) =  \frac{1}{2}\,  \frac{d}{dJ}\, \ln \left[ \frac{\det (J\mu^{-1} O)}{\det_{\bar{\Omega}} 
(J\mu^{-1}O)} \right]_{J=1}  \;.
\]
In addition, we note that,
\begin{eqnarray} 
 \int [D\phi]\, P^{(0)}[\phi]   \int dx \,  \varphi_\Omega (x)[O \varphi_\Omega] (x) =  \int [Df_\Omega]\,  P^{(0)}_\Omega[f_\Omega]  \int dx \,  \varphi_{\Omega} (x)[O \varphi_{\Omega}] (x)  \;, \nonumber 
\label{} 
\end{eqnarray} 
as $\int dx \,  \varphi_{\Omega} (x)[O \varphi_{\Omega}] (x)$ only depends on values of $\phi(x)$ with $x\in \Omega$, through the dependence of $\varphi_\Omega$ on $f_\Omega(x) = \phi(x)|_\Omega $. Then, using a similar expression with $\Omega \leftrightarrow \bar{\Omega}$, we finally get,
\begin{eqnarray} 
\lefteqn{ I(\Omega;\bar{\Omega}) = \frac{1}{2}
\ln \left[ \frac{(\det_\Omega (\mu^{-1}O)\det_{\bar{\Omega}}(\mu^{-1} O)}{\det (\mu^{-1}O)} \right] } 
\nonumber \\
&& ~~~~~~~~ ~~~~~~ -\frac{1}{2}\,  \frac{d}{dJ}\, \ln \left[ \frac{\det_{\Omega} 
(J\mu^{-1}O)\det_{\bar{\Omega}}  
(J\mu^{-1}O)}{ \det (J\mu^{-1} O)} \right]_{J=1}  \;.
\label{ioob}   
\end{eqnarray} 
In section \ref{geomandiv}, we will analyze the geometric and scaling aspects of this representation. In the next section, we shall analyze coherent states and the Fisher information as an interesting measure in Quantum Field Theory.

\section{Probing quantum states}
\label{probe}

For a coherent state, the probability density $P^{(h)}[\phi]$ is given by a field translation of the vacuum state. The field distribution is peaked at a classical solution to the equations of motion,  
\begin{equation}
P^{(h)}[\phi] = P^{(0)}[ \phi -h  ] \makebox[.5in]{,}  (\partial^2_t -\Delta )\, h=0  \;.
\end{equation}
Then, the reduced probability computed by the saddle point method is, 
\begin{eqnarray}
P^{(h)}_\Omega[f_\Omega]  =  \frac{[\det (\mu^{-1} O)]^{1/2}}{[\det_{\bar{\Omega}} (\mu^{-1}O)]^{1/2}} \exp\left[- \int dx \, \bar{\varphi}_\Omega(x) [O \bar{\varphi}_\Omega] (x)\right] \;, 
\label{Pcoh}
\end{eqnarray}
\begin{equation} 
[(-\Delta)^{1/2} \bar{\varphi}_\Omega ](x) = 0 \makebox[.3in]{,} x \in \bar{\Omega}   \makebox[.5in]{,}  \bar{\varphi}_\Omega(x) = f_{\Omega}(x) - h(x)\makebox[.3in]{,}  x \in \Omega \;.
\end{equation} 
In this form, it is clear that the Shannon mutual information for a general coherent state and for the vacuum coincide. The distributions only differ by the change, $f_\Omega \to f_\Omega - h$, $x \in \Omega$, which is irrelevant upon  path-integrating over $f_\Omega$. 
 This also occurs with the entanglement entropy for coherent states, as reported in Ref. \cite{37}.  In the next subsection, we shall consider the Fisher information as a means to get a space-dependent information measure in relativistic QFT, where the space points are field labels and not fluctuating quantum observables.
This measure will display a nontrivial dependence on coherent states.

\subsection{Fisher information in QFT}

Given a wave functional $\Psi[\phi]$, let us consider the reduced probability $P_{\Omega_u}[f_{\Omega_u}]$ of $P[\phi]$ with respect to a ``probe'' $ \Omega_u$, namely, a family of regions obtained from $\Omega$ by a translation with $u \in \mathbb{R}^n$.  Note that $f_{\Omega_u} (x)$ is a field shape only defined inside $\Omega_u$. Then, the concept of analyzing the probability density for the same shape on different regions amounts to considering $f_{\Omega_u} (x) = f_\Omega (x-u)$, where $f_\Omega(x)$ is defined on $\Omega$. This permits to introduce the relative entropy to compare distributions between reduced field variables on different regions,
\begin{eqnarray}
D(u;u') = \int [Df_\Omega] \, P_{\Omega_u}[f_{\Omega_u}] ( \ln P_{\Omega_u}[f_{\Omega_u}] - \ln P_{\Omega_{u'}}[f_{\Omega_{u'}}] )
\label{dive}
\end{eqnarray}
and then compute the Fisher information \cite{fisher}, which gives its infinitesimal variation $\delta D = \frac{1}{2} \, \delta u_i I_{ij} \delta u_j$, and is related with the information that the reduced probability carries about the parameter $u$, 
\[
I_{ij} (u) = \left. \partial^{u'}_i \partial^{u'}_j\,  D(u;u') \right|_{u=u'} \makebox[.5in]{,} \partial^{u'}_i =\frac{\partial}{\partial u'_i}  \;.
\]
Let us obtain this measure for a coherent state. The reduced probability with respect to $\Omega$ depends on
(cf. eq. (\ref{Pcoh}))
\[
\int dx \, \bar{\varphi}_\Omega(x) [O \bar{\varphi}_\Omega] (x) =  \int_\Omega dx \, (f_\Omega(x) -
h(x)) [O \bar{\varphi}_\Omega] (x) \;.
\]
\begin{equation}
\bar{\varphi}_\Omega(x)  = \left\{ \begin{array}{ll}
          f_\Omega(x) - h(x) & \mbox{if $x \in \Omega $} \\
       \int_{\Omega} dy\, \mathscr{P}(x,y) (f_\Omega(y) -h(y))& \mbox{if $x \in \bar{\Omega}$}\;. \end{array} \right. 
\end{equation}  
$\mathscr{P}(y,z)$ is the Poisson kernel associated with $\Omega$ (for a ball, see eq. (\ref{Pik})).
 When $x \in \Omega$, 
\begin{eqnarray}
\lefteqn{  [O \bar{\varphi}_\Omega](x) 
=  {\cal C}_1 \int_{\Omega} dy  \, \frac{(f_\Omega (x)- h(x))- (f_\Omega (y)-h(y) )}{|x-y|^{d+1}} } \nonumber \\
&& +  {\cal C}_1 \int_{\bar{\Omega}} dy  \, \frac{(f_\Omega(x)- h(x)-\bar{\varphi}_\Omega( y))}{|x-y|^{d+1}}   
 \;.  \nonumber
\end{eqnarray}
Now, when reducing with respect to $\Omega_u$, the probability in eq. \eqref{Pcoh} will contain 
a functional determinant $\det_{\bar{\Omega}_u} (\mu^{-1}O)$, which  does not depend on $u$, and the Poisson problem in $\Omega_u$ will be solved by  $ \mathscr{P}(y-u,z-u)$.  Then, for a given field shape, the probability distribution in the translated region $P^{(h)}_{\Omega_u}[f_{\Omega_u}]$ is given by the probability distribution in $\Omega$, $P^{(h_u)}_{\Omega}[f_{\Omega}]$ computed for a coherent state wave functional distributed around $ h_u(x) = h(x+u)$. Of course, for the ground state, which corresponds to $h(x)=0$, the probability distribution  is $u$-independent, as expected. In particular, for a constant shape $f_\Omega \equiv \psi$, it is given by,
\begin{eqnarray}
\frac{(\det (\mu^{-1}O))^{1/2}}{(\det_{\bar{ \Omega} }  (\mu^{-1}O))^{1/2}} \, e^{- {\cal C}_1\, \psi^2 \int_{\Omega} dx \, \int_{\bar{\Omega}} dy  \, \left(1-\int_{\Omega} dz\, \mathscr{P}(y ,z )\right) 
  |x-y|^{-d-1}  }  \;. \nonumber 
\end{eqnarray} 

For a general coherent state, eq. (\ref{dive}) gives,
\begin{eqnarray}
D(u;u') = \int [Df_\Omega] \, P^{(h_u)}_{\Omega}[f_{\Omega}] \left( \ln P^{(h_u)}_{\Omega}[f_{\Omega}]- \ln P^{(h_{u'})}_{\Omega}[f_{\Omega}]  \right)  \;.
\end{eqnarray}  
Changing variables $f_\Omega  \to f_\Omega +  h_u$,  $P^{(h_u)}_{\Omega}[f_{\Omega}]$ becomes $  P^{(0)}_\Omega[f_\Omega]$, that is, the reduced probability for the ground state. Then, we arrive at,
\[ 
D(u;u') =  - \int [Df_\Omega] P^{(0)}_\Omega[f_\Omega] \ln P^{(h_{u'})}_\Omega[f_\Omega + h_u] +  \dots
\]
where the dots represent $u'$-independent terms. In addition, the average of 
quadratic and linear terms in $f_\Omega$ are $u'$-independent and vanishing, respectively.
Using the normalization condition, this leads to, 
\[ 
D(u;u') =    \int_\Omega dx \, \chi(x) [O \chi ] (x)  +  \dots  \;, 
\]
where the dots now include additional $u'$-independent terms originated from the functional determinants, and
\begin{eqnarray}
 [O \chi](x) 
=  {\cal C}_1 \int_{\Omega} dy  \, \frac{ \delta h(x)- \delta h(y)}{|x-y|^{d+1}}  +  {\cal C}_1 \int_{\bar{\Omega}} dy  \, \frac{ \delta h(x) - \chi( y)}{|x-y|^{d+1}}   
 \;,  \nonumber
\end{eqnarray}
\begin{equation}
\chi (x)  = \left\{ \begin{array}{ll}
          \delta h (x) & \mbox{if $x \in \Omega $} \\
       \int_{\Omega} dy\, \mathscr{P}(x,y) \delta h (y) & \mbox{if $x \in \bar{\Omega}$}\;, \end{array} \right. 
\end{equation} 
$\delta h(x) = h_{u'}(x) -h_u(x) $. The Fisher information matrix is then,
\[ 
I_{ij}(u) = 2\int_\Omega dx \, f_i(x,u) [Of_j](x,u)  \;,
\]
\begin{eqnarray}
 [O f_j](x,u) 
=  {\cal C}_1 \int_{\Omega} dy  \, \frac{ \partial_j h_u(x)- \partial_j h_u(y)}{|x-y|^{d+1}}  +  {\cal C}_1 \int_{\bar{\Omega}} dy  \, \frac{ \partial_j h_u(x) - f_j( y,u)}{|x-y|^{d+1}}    
 \;,  \nonumber
\end{eqnarray}
\begin{equation}
f_i (x,u)  = \left\{ \begin{array}{ll}
          \partial_i h_u (x)& \mbox{if $x \in \Omega $} \\
       \int_{\Omega} dy\, \mathscr{P}(x,y)\, \partial_i h_u(y) & \mbox{if $x \in \bar{\Omega}$}\;. \end{array} \right. 
\end{equation} 
where the derivatives are taken with respect to $u^i$. In particular, 
considering a small ball centered at $x=0$, with radius much smaller than the scale of spatial variations of the classical solution, we get,
\[ 
I_{ij}(u) \approx 2 v\,  [Of](0) \, T_{ij} \makebox[.5in]{,}
f (x)  = \left\{ \begin{array}{ll}
         1& \mbox{if $x \in \Omega $} \\
       \int_{\Omega} dy\, \mathscr{P}(x,y) & \mbox{if $x \in \bar{\Omega}$}\;, \end{array} \right. 
\]
where $v$ is the volume of $\Omega$ and $T_{ij} =  \partial_i h(u) \partial_j h(u)$ is the stress-tensor for the classical solution $h(u)$. It is also interesting to note that the reduced probabilities can also be compared at different times, leading to a relative entropy $D(u,t;u',t')$. For coherent states, the time dependence is present in the mean field value $h(x,t)$, and the generalized Fisher information matrix becomes 
$I_{\mu \nu}(u,t) \approx 2 v\,  [Of](0) \, T_{\mu \nu}(u,t)$, with $T_{\mu\nu}(u,t) =  \partial_\mu h(u,t)  \partial_\nu h(u,t)$.

\section{Geometric contributions}
\label{geomandiv}

To compute the Shannon mutual information in eq. \eqref{ioob}, we need a framework to deal with functional determinants. For example, we can use the heat kernel definition \cite{Hawking} (see also \cite{12}, and references therein).  In this case, for an operator $A$ defined on some region $M$, with conditions on $\partial M$ (local case) or on $\bar{M}$ (nonlocal case), the functional determinant is obtained from, 
\begin{equation}
\ln {\det}_{M} \left(\mu^{-n} A \right)  \equiv  -\lim_{\epsilon \to 0} \int_{\epsilon}^\infty dt \, t^{-1}\,  Y_{A,\mu}(t) \;,
\label{defh}
\end{equation} 
where $ Y_{A,\mu}(t)$ is the trace of the heat kernel $F_{A,\mu}(x,y,t)$,
\begin{equation}
Y_{A,\mu}(t) = \int_M dx\, \sqrt{g}\ F_{A,\mu}(x,x,t) \;,
\label{tY}
\end{equation}
\begin{equation}
\frac{d}{dt}F_{A,\mu}(x,y,t) + \mu^{-n} A\, F_{A,\mu}(x,y,t)=0 
\makebox[.5in]{,}
F_{A,\mu}(x,y,0) = \delta(x-y) 
\label{hk}
\end{equation}
($n$ is the mass dimension of $A$, while $g$ is the determinant of the metric on $M$).
The heat kernel satisfies,  in the $x$ and $y$ variables, the same conditions needed to compute the determinants. It represents the diffusion of a unit quantity of heat placed at $y$ in $t=0$.
Now, taking the heat equation for $ O= (-\Delta )^{\frac{1}{2}}$ defined on $M$, with Dirichlet conditions on $\bar{M}$, 
\begin{equation} 
\frac{d}{dt}F_{O,\mu}(x,y,t) + \mu^{-1} O\,  F_{O,\mu}(x,y,t) =0  \;,
\label{heattt}
\end{equation}
applying $O$, and using the fractional equation again, we get, 
\begin{equation}
-\frac{\partial^2}{\partial t^2}F_{O,\mu}(x,y,t) - \mu^{-2} \Delta \, F_{O,\mu}(x,y,t)  =0 \;,
\end{equation}
which is solved by,
\begin{equation}
F_{O,\mu}(x,y,t) = \sum_i \Phi_i(x)\Phi_i(y)\, e^{- \left( \frac{\lambda_i}{\mu^2}\right)^{\frac{1}{2}}\, t}  \;, 
\label{soloc}
\end{equation}
where $\Phi_i$ are the eigenfunctions of $-\Delta$ and $\lambda_i$ the respective eigenvalues. Then,
\[
 Y_{O,\mu}(t) = \sum_i e^{- \left( \frac{\lambda_i}{\mu^2}\right)^{\frac{1}{2}}\, t} \;,
\]
which after replacing in eq. (\ref{defh}) makes it clear that the following natural relation can be used,
\begin{eqnarray}
\ln {\det}_{M} \left( \mu^{-1}O \right) & = & \frac{1}{2} \sum_i \ln \frac{\lambda_i}{\mu^2} \nonumber \\
& = &  \frac{1}{2} \ln {\det}_M \left[\mu^{-2} O^2\right]
  \;,
\end{eqnarray} 
to get,
\begin{eqnarray} 
 I(\Omega;\bar{\Omega}) = \frac{F}{2}  -\frac{1}{4}\,  \frac{d}{dJ}\, \ln \left[ \frac{\det_{\Omega} 
 \left[J^2\mu^{-2} O^2 \right]\det_{\bar{\Omega}}  
\left[J^2\mu^{-2} O^2 \right]}{ \det \left[J^2\mu^{-2} O^2 \right]} \right]_{J=1}  \;, 
\label{info}
\end{eqnarray} 
\begin{equation}
F =  \frac{1}{2} 
\ln \left[ \frac{(\det_\Omega \left[\mu^{-2} O^2 \right]\det_{\bar{\Omega}}\left[\mu^{-2} O^2 \right]}{\det \left[\mu^{-2} O^2 \right]} \right] \;.
\label{class} 
\end{equation} 
This is a definite relation between the Shannon mutual information $ I(\Omega;\bar{\Omega})$ (cf. eq. \eqref{IOm})  and $F$, the continuum version of the classical mutual correlation discussed in Ref. \cite{Cramer}. For the computation of both quantities in one dimensional field theories, see Ref. \cite{22}. As we will see, the second term in eq. \eqref{info} does not modify the logarithmic divergence nor the scaling properties of the first. 

\subsection{Divergent terms}

To understand the structure of the divergences of the mutual information, we can rely on the heat kernel method for $A=-\Delta$. In fact, for the general form 
$A = -\nabla_a\nabla^a + c\, R$, where $ \nabla_a $ is the covariant derivative and $R$ is the curvature scalar, 
the small $t$-expansion is \cite{4}
\begin{equation}
Y_{A,\mu}(t) \approx \frac{(\mu^2)^{\frac{d}{2}}}{(4\pi t)^{\frac{d}{2}}}\sum_{j=0}^\infty C^M_j \left(\frac{t}{\mu^2}\right)^{j/2}  \;,
\label{expheat}
\end{equation}
where the Seeley-De Witt coefficients $C^M_j$ depend on geometrical properties of $M$ and its border. In addition, it is known that  $Y_{A,\mu}(t)$ decays exponentialy for $t\rightarrow\infty$ \cite{12}. Therefore, it is easy to identify the divergent terms of
\begin{equation}
 \ln {\det}_M [\mu^{-2} (-\Delta)]  = -\lim_{\epsilon\to 0} \int_{\epsilon}^{\infty} dt\, t^{-1}Y_O(t) 
\end{equation}  
in an $\epsilon$-expansion \cite{Hawking,12},
\begin{eqnarray}
 \ln {\det}_M [\mu^{-2} (-\Delta)] =   \frac{2}{(4\pi)^{\frac{d}{2}}}  \sum_{j=0}^{d-1} C^M_j \, \frac{ \delta^{j-d} }{j-d}  + \frac{2}{(4\pi)^{\frac{d}{2}}} \, C^M_d \ln \left( \frac{\delta}{a} \right) + \dots
\label{geomdivpart}
\end{eqnarray}
where $\delta = \sqrt{\epsilon} /\mu$, $a = \sqrt{t_0}/\mu $ and $t_0$ is such that $Y_{A, \mu}(t)$ is well described by eq. \eqref{expheat} for $t < t_0$. For the second term in eq. \eqref{info}, the divergent part of $\ln {\det}_M [J^2 \mu^{-2} (-\Delta)] $ is simply obtained by replacing $\delta \to J \delta$, $a \to J a $ in eq. \eqref{geomdivpart}. Therefore, in $d$ dimensions, the geometric and divergent part of the Shannon mutual information is,
\begin{eqnarray}
 I_d(\Omega;\bar{\Omega})|_{\rm div} =   \frac{1}{2(4\pi)^{\frac{d}{2}}}  \sum_{j=0}^{d-1} \left( C^{\Omega}_j + C^{\bar{\Omega}}_j - C_j \right) \, \frac{1+ d -j  }{j-d}\,  \delta^{j-d} \nonumber \\
 +\frac{1}{2(4\pi)^{\frac{d}{2}}}\,  \left( C^{\Omega}_d + C^{\bar{\Omega}}_d - C_d \right) \, \ln \left( \frac{\delta}{a} \right) \;,
  \label{div-i}
\end{eqnarray}
where $C_j$ denotes the coefficients for the whole space $\Omega \cup \bar{\Omega}$. More precisely (\cite{12}, \cite{4}, \cite{5},  and Refs. therein),
\begin{equation}
C^M_j = A^M_j+ B^M_j = \int_M dx\, \ \sqrt{g}\ a^M_j + \int_{\partial M} dy\, \ \sqrt{\hat{g}}\ b^M_j \;,
\label{coef}
\end{equation}
where $a^M_j$ depends on the Riemann tensor, its covariant derivatives, and is independent of the boundary condition, while $b^M_j$ depends on the type of boundary condition (Dirichlet, Neumann, Robin, etc), is a polynomial in the metric, the boundary's normal vector and its covariant derivates ($\hat{g}$ is the determinant of the induced metric $\hat{R}$ on $\partial M$). Explicit values were obtained in \cite{12}, relying on Refs. \cite{Hawking}, \cite{Seeley}. For Dirichlet boundary conditions, the initial values are, $a^M_0=1$, $b_0=0$, $a^M_1=0$,  $a^M_1=0$, $b^M_1=-\frac{\sqrt{\pi}}{2}$. They imply 
\[
C^M_0 = \int_M dx\, \sqrt{g} = {\rm Vol}_M \makebox[.5in]{,} C^M_1 = -\frac{\sqrt{\pi}}{2} \int_{\partial M} dy\, \sqrt{\hat{g}} = -\frac{\sqrt{\pi}}{2}\, {\rm A}_{\partial M}\;,
\]  
so that the volumetric term in eq. \eqref{div-i}, associated with $j=0$, is canceled. 

In $d=1$, the area is obtained as usual, taking the $d\to 1$ limit of  $2\pi^{\frac{d}{2}}/\Gamma(\frac{d}{2})$, which is  the area of the border of a $d$-dimensional ball. That is, ${\rm A}_{\partial M} \to 2$ in the case of a line segment, or, more generally, $A_{\partial M} \to N$, where $N$ is the number of points in the border of $M$. Therefore, in this case, the logarithmic divergence is
\begin{equation}
-\frac{1}{8}\left(N_{\partial\Omega}+N_{\partial\bar\Omega} - N_S\right)\ln\left(\frac{\delta}{a}\right) =
-\frac{\cal{A}}{8} \, \ln\left(\frac{\delta}{a}\right) \;,
\end{equation}
where $N_{\partial\Omega}$, $N_{\partial\bar\Omega}$ and $N_S$ are the number of points on the border of $\partial\Omega$, $\partial\bar{\Omega}$ and on the physical space, respectively, and $\cal{A}$ is the number of points between $\Omega$ and its complement. This $\cal{A}$-dependence is expected to occur for the entanglement entropy as well (see Ref. \cite{Cardy}).

In general, $a_j$-coefficients do not contribute to the divergent part. They vanish for odd $j$, and for 
even $j$, the integral over $\Omega$ and $\bar{\Omega}$ minus that over $\Omega \cup \bar{\Omega}$ is always zero. Then, the divergences only depend on  properties at the boundaries. 
For $d \geq 2$, 
\begin{eqnarray}
I_d(\Omega;\bar{\Omega})|_{\rm div} =  \frac{\sqrt{\pi}}{4(4\pi)^{\frac{d}{2}}}   \frac{d}{d-1}\,  \delta^{1-d} \,  {\rm A}_{\partial M} + K_d \nonumber \\
  +  \frac{1}{2(4\pi)^{\frac{d}{2}}}\, \left( C^{\Omega}_d + C^{\bar{\Omega}}_d - C_d \right) \, \ln \left( \frac{\delta}{a} \right) \;,
  \label{div-m2}
\end{eqnarray}
where $K_2 = 0$ and for $d\geq 3$,
\begin{eqnarray}
K_d =  \frac{1}{2(4\pi)^{\frac{d}{2}}}  \sum_{j=2}^{d-1} \left( C^{\Omega}_j + C^{\bar{\Omega}}_j - C_j \right) \, \frac{1+ d -j  }{j-d}\,  \delta^{j-d}  \;.
\end{eqnarray}
The divergent terms have the same form expected for the entanglement entropy (eq. \eqref{arealaw}), including the famous area law. 
For even $j$, the coefficients $b^{\Omega}_j$ are odd in the boundary's normal vector, thus implying 
\[
C^{\Omega}_j + C^{\bar{\Omega}}_j - C_j = 0\;.
\]  
In particular, for even spatial dimensions, the coefficient of the logarithmic divergence is zero. For odd $j$
 \[
C^{\Omega}_j + C^{\bar{\Omega}}_j - C_j = 2C^\Omega_j\;,
\]
and the logarithmic divergence for odd spatial dimensions is,
\begin{equation}
I_d(\Omega;\bar{\Omega})|_{\rm log}  = \frac{C_d^\Omega}{(4\pi)^\frac{d}{2}} \, \ln \left( \frac{\delta}{a} \right) \;.
\label{UAUWW}
\end{equation}  
An algorithm to compute the coefficients $C_j^\Omega$ in any dimension and geometry was developed in \cite{5}. For a ball in two, three and four dimensional Euclidean space, the coefficients $C_j^\Omega$ can be found in \cite{sphere}, \cite{sphere1}. They lead to,
\[
I_2(B;\bar{B})|_{\rm div} =  \frac{\sqrt{\pi}}{4} \,  \delta^{-1} \,  {\rm r}  \makebox[.5in]{,}
I_3(B;\bar{B})|_{\rm div} =  \frac{3}{16} \,  \delta^{-2} \,  {\rm r^2} -\frac{1}{48}\, \ln \left( \frac{\delta}{a} \right) \;,
\] 
\[
I_4(B;\bar{B})|_{\rm div} =  \frac{\sqrt{\pi}}{24} \,  \delta^{-3} \,  {\rm r^3} +\frac{11\sqrt{\pi}}{256}\,\delta^{-1}\,{\rm r} \;.
\] \label{div-m3}

\subsection{Scaling properties}

When the spectrum of the operator $O$ appearing in the Gaussian wave functional is known, the full expression for $I(\Omega,\bar{\Omega})$ can be obtained via zeta function. This is not a procedure that regularizes the divergences, permitting to ``see'' and then eliminate them by means of a renormalization, but one that already gives finite answers. For $O= (-\Delta)^{\frac{1}{2}}$, following similar steps to those given from eq. \eqref{heattt} to \eqref{soloc}, we can relate the eigenvalues of $O$ and $O^2$ to get eq. (\ref{info}) with,
\begin{equation}
\ln  {\det}_M  (  \mu^{-2} O^2) \equiv -\zeta'_{M}(0)   \makebox[.5in]{,}
\zeta_{M}(s) = \sum_i \left(\frac{\lambda_i}{\mu^2}\right)^{-s} \;, 
\end{equation}
where $\zeta_{M}(s)$ is the analytic continuation of the function $\sum_i \left(\frac{\lambda_i}{\mu^2}\right)^{-s}$ defined on its region of convergence. 
For a line segment of length $l$, with Dirichlet boundary conditions, the eigenvalues of the Laplacian are  $\frac{q^2\pi^2}{l^2}$, with $q \in \mathbb{N} - \{ 0 \}$. Then,
\begin{equation}
\zeta_M(s)=\left(\frac{\pi^2}{\mu^2l^2}\right)^{-s}\sum_{q=1}^{\infty} q^{-2s}=\left(\frac{\pi^2}{\mu^2l^2}\right)^{-s}\zeta(2s)  \;,\label{di}
\end{equation}
($\zeta(\cdot )$ is the Riemann zeta function) and  the logarithm of the determinant is, 
\begin{equation}
-\zeta'_M(0) =\ln\left(\frac{\pi^2}{\mu^2 l^2}\right)\zeta(0) - 2\zeta'(0) = \ln(2\pi) + \ln\left(\frac{l\mu}{\pi}\right) =  \ln\left(2 l\mu\right)  \;.
\label{exemplinho} 
\end{equation} 
Thus, for a system of size $L$ divided into two adjacent parts, with size $l$ and $L-l$, respectively, the mutual information is,
\begin{equation}
I(\Omega,\bar{\Omega})= \frac{F}{2}  +\frac{1}{4} \makebox[.5in],{} F = \frac{1}{2}\ln\left(\frac{2\mu\, l(L-l)}{L}\right) \;.
\label{direct}
\end{equation}
This (continuum) classical mutual correlation $F$ agrees with the result in Ref. \cite{22}, based on a lattice regularization, 
\[
\frac{1}{2}\ln\left(\frac{ n(N-n)}{N}\right)  \;.
\]
where $n$, $N$ is the number of lattice points in each region.  

For the massive case with Dirichlet boundary conditions, we can use (see Ref. \cite{ratio}
and references therein), 
\begin{equation}
\frac{\det\left(-\frac{1}{\mu^2}\frac{d^2}{dx^2} + \frac{1}{\mu^2}R_1(x)\right)}{\det\left(-\frac{1}{\mu^2}\frac{d^2}{dx^2}+\frac{1}{\mu^2}R_2(x)\right)} = \frac{\det\left(-\frac{d^2}{dx^2} + R_1(x)\right)}{\det\left(-\frac{d^2}{dx^2}+R_2(x)\right)} =\frac{y_1(l)}{y_2(l)}\;,
\end{equation}
where $y_i(x)$ is the unique solution of
\begin{equation}
\left(-\frac{d^2}{dx^2}+R_i(x)\right) y_i(x)=0  \makebox[.5in]{,} y_i(0) = 0 \makebox[.5in]{,} y'_i(0)=1\;.
\end{equation}
In addition, the massive and massles zeta functions coincide at $s=0$ (see Ref. \cite{12}). Therefore,
\begin{equation}
\det\left(-\frac{1}{\mu^2}\frac{d^2}{dx^2} + \frac{m^2}{\mu^2}\right)=\frac{\sinh(ml)}{ml}\det\left(-\frac{1}{\mu^2}\frac{d^2}{dx^2}\right) \;,
\end{equation}
\begin{equation}
I_m(\Omega,\bar{\Omega}) =\frac{1}{4} \ln\left(\frac{2\mu \sinh(ml) \sinh(m(L-l))}{m \sinh(mL)}\right)+\frac{1}{4} \;.
\end{equation}
For large $m$, the $l$ and $L$ dependence disappears, meaning that the field variables become independent. In fact, the mutual information can be set to zero  in this limit by subtracting the asymptotic behavior $I_m(\Omega,\bar{\Omega}) \to \frac{1}{4} \ln\left(\frac{2\mu }{m }\right) +\frac{1}{4}$, which has a dependence on the correlation length $1/m$ similar to that observed for the entanglemement entropy \cite{Cardy}. 

In any dimension, if $A$ is a second order differential operator, we can use the property \cite{12},
\begin{equation}
\ln {\det}_M  \left(J^2 \mu^{-2}A\right) = \ln {\det}_M \left(\mu^{-2} A\right) + \zeta_{M}(0) \ln J^2 \;,
\label{scaling}
\end{equation} 
\begin{equation}
\zeta_{M}(0) = \frac{1}{(4\pi)^{d/2}} \, C_d^M  \;,
\label{UAUW}
\end{equation} 
to conclude that the zeta-regularized Shannon mutual information is,
\[
I(\Omega;\bar{\Omega})  = -\frac{1}{4}\, [\zeta'_{\Omega}(0) + \zeta'_{\bar{\Omega}}(0) -\zeta'_S(0) ] -\frac{1}{2}\, [ 
 \zeta_{\Omega}(0) +  \zeta_{\bar{\Omega}}(0) -  \zeta_{S}(0) ] \;. 
\]
Note that this expression could be extended for a general curved static spacetime, when the vacuum wave functional is Gaussian. 
Now, for the massless case $A=-\Delta$ on an Euclidean space, when changing the length scales by a factor $\kappa$, the eigenvalues get scaled by ${\kappa}^{-2}$. Using  eq. (\ref{scaling}) with $J = \kappa^{-1} $,
\begin{equation}
\ln {\det}_M  \left(\mu^{-2}A\right) \to  \ln {\det}_M \left(\mu^{-2} A\right) - 2\zeta_{M}(0) \ln \kappa \;,
\end{equation}  
\[
I(\Omega;\bar{\Omega})  \to I(\Omega;\bar{\Omega})  - \frac{1}{2(4\pi)^{d/2}} \, (C_d^\Omega + C_d^{\bar{\Omega}} -C_d^S )   \ln \kappa   \;.
\]
Then, the following remarks are in order. First, the Seeley-De Witt coefficients in the logarithmic divergence in eq.  
(\ref{UAUWW}), when using the heat-kernel, coincide with the coefficients that dictate how the Shannon mutual information behaves under scale transformations.
Second, for simple geometries, the scaling law of the functional determinant strongly restricts the form of the mutual information. 
When $M$ is a segment of size $l$, the calculation of $\ln {\det}_M  \left(\mu^{-2}A\right)$ must contain a logarithmic term,
\begin{eqnarray}
 - 2\zeta_{M}(0) \ln \mu l = - \frac{1}{\sqrt{\pi}} \, C_1^{M} \ln \mu l = \frac{1}{2} \, {\rm A}_{\partial M} \ln \mu l =  \ln \mu l
\end{eqnarray}
plus contributions that do not scale. As the only variable that scales is $l$, this contribution must be a constant. Then, the Shannon mutual information for a partitioned segment must be, 
\[  
I(\Omega, \bar{\Omega}) = \frac{1}{4}\, \ln \frac{\mu l (L-l)}{L} + {\rm const.}  \;,
\]
in accordance with the direct calculation in eq. (\ref{direct}). Similarly,
for the mutual information in $d$ Euclidean dimensions, reducing the probabilities with respect to a ball $\Omega = B_r$ of radius $r$, and the field theory defined on the ball $B_R$ ($R > r$), 
\[  
\ln {\det}_{B_r}  \left(\mu^{-2}A\right) = - \frac{1}{(4\pi)^{d/2}} \, 2 C_d^{B_r} \ln \mu r + {\rm const.}
\]
A direct calculation of this determinant can be found in ref. \cite{outraeli}.
For the region $\bar{\Omega} = B_R - B_r$, we must have, 
\[
\ln {\det}_{\bar{\Omega}}  \left(\mu^{-2}A\right) =  - \frac{1}{(4\pi)^{d/2}} \, 2  C_d^{\bar{\Omega}} \ln \mu (R-r)  + F(r/R) \;.
\]
In addition, in an Euclidean space, the $a_j$-coefficients are always zero, while
\begin{eqnarray}
b_2 = \frac{\chi}{3} \makebox[.5in]{,} b_3 = \frac{\sqrt{\pi}}{192}\, (9\chi^2 -6\chi_{ij}\chi_{ij}- 
16 \hat{R})  \;,
\end{eqnarray}  
see Ref. \cite{12}  ($\chi_{ij} = - \partial_i n_j$ , $\chi = \chi_{ii}$ and $\hat{R}$ is the curvature at the boundary).
Then, from eq. \eqref{coef}, in two and three dimensions, the logarithmic contributions to the mutual information in $d=2,3$ turn out to be,
\[
I_2(\Omega, \bar{\Omega}) = -\frac{1}{12} \ln r/R  + \dots \;,
\]
\[
I_3(\Omega, \bar{\Omega}) = \frac{1}{96} \ln \left[ \mu^2 \frac{r}{R} (R-r)^2 \right]  + \dots \;.
\] \label{div-m4}

\section{Conclusions}
\label{conc}

In this work, we presented a direct calculation of the Shannon mutual information in the continuum, for a massless scalar field and a general differentiable manifold. Initially, we showed how the nonlocal Poisson problem on a given region is intimately related with the reduced probability of getting a given field shape on its complement. The link is established through the saddle-point uniqueness condition when evaluating Gaussian path-integrals that involve nonlocal kernels. These procedures are essential to account for the correlations between in and out modes, which are expected to be manifested in the mutual information and related quantities.
Following them, we were able to compute the reduced probability for a $d$-dimensional ball, as well as a simple expression for the Shannon mutual information with respect to a general region. This measure turned out to be simply related with the classical mutual information discussed in Ref. \cite{22}, up to a factor $1/2$ and a term that does not modify logarithmic divergences nor scaling properties. Furthermore, the result for a general coherent state coincides with that for the vacuum, a property that was also observed in the case of the entanglement entropy \cite{37}. We also studied the Fisher information associated with a ``probe'' $\Omega_u$, where $u \in \mathbb{R}^d$ is the translation parameter of a region $\Omega$. This is an interesting way of defining, in a relativistic Quantum Field Theory, an information measure associated with points in the physical space, although they are labels and not quantum variables.  

Next, to obtain explicit results about the Shannon mutual information $I(\Omega;\bar{\Omega})$, we had to rely on different ways to deal with functional determinants. For example, using the heat-kernel regularization,  divergences can be obtained in terms of the Seeley-De Witt coefficients, which codify information about the geometry of $\Omega$. In any dimension, the divergences are  dominated by an area law, 
while the coefficient of the logarithmic divergence is related to the behavior of $I(\Omega;\bar{\Omega})$
under scale transformations. The latter property is better understood by using the zeta-function determination of functional determinants. The logarithmic  coefficient is zero in even dimensions, a result that agrees with  R\'enyi-n for a $d$-dimensional ball \cite{2}. It is also consistent with a special case calculated  in Ref. \cite{22}, for massive and massless scalar fields in $d=1$. Finally, the scaling properties also permitted to obtain  the logarithmic dependence of the Shannon mutual information on the system size, when $\Omega$ is a ball in two or three dimensions.

\section*{Acknowledgements}

We would like to thank C. D. Fosco, M. A. Rajabpour and M. Moriconi for useful discussions.
The Conselho Nacional de Desenvolvimento Cient\'{\i}fico e Tecnol\'{o}gico (CNPq), the Coordena\c c\~ao de Aperfei\c coamento de Pessoal de N\'{\i}vel Superior (CAPES), and FAPERJ are acknowledged for the financial support.

\end{document}